\documentclass[twocolumn,showpacs,aps,pra]{revtex4}
\usepackage{epsfig}

\begin{document}
\title{Measuring electric fields from surface contaminants with neutral atoms}
\author{J.~M. Obrecht$^{\star}$, R.~J. Wild, E.~A. Cornell$^{\dag}$}
\affiliation{JILA, National Institute of Standards and Technology and University of Colorado, Boulder, Colorado 80309-0440, USA
\\and Department of Physics, University of Colorado, Boulder, Colorado 80309-0390, USA}
\date{\today}

\begin{abstract}

In this paper we demonstrate a technique of utilizing magnetically trapped neutral $^{87}$Rb atoms to measure the magnitude and direction of
stray electric fields emanating from surface contaminants. We apply an alternating external electric field that adds to (or subtracts from) the
stray field in such a way as to resonantly drive the trapped atoms into a mechanical dipole oscillation.  The growth rate of the oscillation's
amplitude provides information about the magnitude and sign of the stray field gradient.  Using this measurement technique, we are able to
reconstruct the vector electric field produced by surface contaminants.  In addition, we can accurately measure the electric fields generated
from adsorbed atoms purposely placed onto the surface and account for their systematic effects, which can plague a precision surface-force
measurement. We show that baking the substrate can reduce the electric fields emanating from adsorbate, and that the mechanism for reduction is
likely surface diffusion, not desorption.

\end{abstract}


\maketitle

\section{Introduction}

The advent of cold-atom technology has brought to light a significant amount of knowledge of the physical world, and has also contributed
significantly to technology such as time standards and global synchrony. Many precision measurements and experimental realizations have taken
advantage of the extremely slow nature of ultracold atoms, which has resulted in such phenomena as Bose-Einstein condensation (BEC), quantized
vortices, ultracold molecules, and atomic parallels to laser optics, to name a few. Recently, the scalability and high level of precision of
ultracold atomic systems have led to an increase in their use as precision tools to measure forces and fields at both large~\cite{gravkas1,
blochingo, gravtino, blochtino, gravkas2} and small~\cite{aspectqr, hinds, daveJLTP, jeffPRA, davePRA, vdwkronin, magschr, ourPRL} length
scales.

What makes ultracold atomic systems so attractive for precision use is the purity of the actual measurement device, the atoms. One may think of
a collection of ultracold atoms as being a large sample of extremely small, yet sensitive, devices that connect to the outside world through
trapping fields and narrow linewidth lasers only, with no physical contacts to transfer heat or mechanical and electrical noise. The sensitivity
of the device can be tuned by selecting the correct atomic species and desired internal electronic state to meet one's specifications. An atomic
ensemble therefore is a tunable system, whose sensitivity (or insensitivity) is well characterized and changeable at the microsecond time scale.

In this paper, we further develop a method of measuring small electric fields near bulk materials with a magnetically trapped BEC of $^{87}$Rb
atoms~\cite{jeffPRA,davePRA}. As a test of our ability to measure these electric fields, several clouds of ultracold atoms were purposely
adsorbed onto a surface to generate a sizeable field.  By measuring the strength of the fields in all three spatial directions, we are able to
fully account for the resulting systematic frequency shifts of mechanical dipole oscillations, such as those reported in other
experiments~\cite{davePRA,jeffPRA,ourPRL}, and estimate the dipole moment per atom adsorbed onto the surface. In addition, the ability of our
magnetic trap to translate along the surface of a bulk substrate allows us to measure electric fields at various surface locations. From these
measurements we can fully reconstruct a three-dimensional vector plot of the electric fields that emanate from the surface, with micron-scale
resolution of the field. Lastly, we investigate the ability to reduce the strength of stray electric fields by diffusing adsorbates across the
surface with heat.

\section{Apparatus}

\begin{figure}
\leavevmode \epsfxsize=3.3in \epsffile{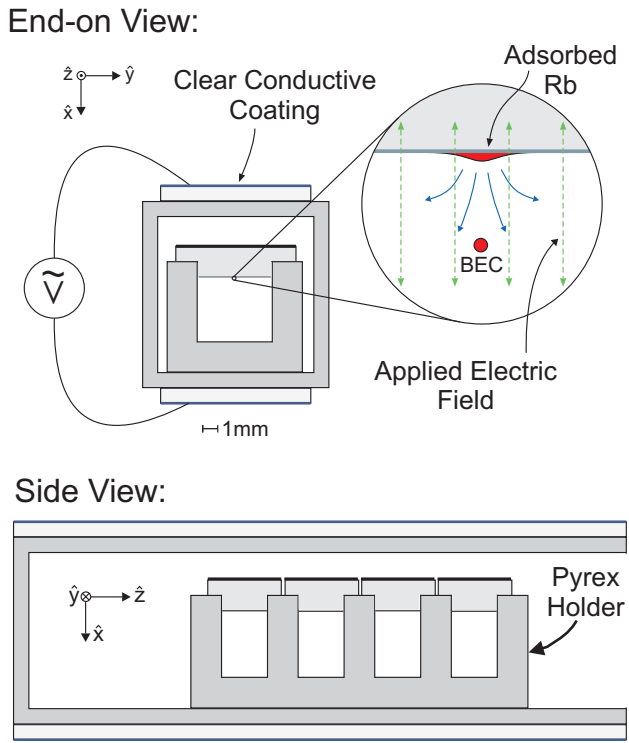}\caption{\label{fig:cell} (Color online). A schematic end-on view (top) and side view (bottom) of
our science cell. Above and below the pyrex cell are the glass plates, coated with a conductive indium-tin-oxide film, which provide the applied
electric field. The BEC, positioned several microns from the metal-coated substrate, sees the oscillating electric field (dashed green arrows)
from the plates and the static electric field (solid blue arrows) from the adsorbed atoms.  The adatoms are illustrated as a mound of material
on the surface, but in reality they have no spatial extent along $\hat{x}$. }
\end{figure}

We use a two-chamber vacuum system to prepare our cloud of ultracold atoms~\cite{Heather2003}. The atoms are loaded from a magneto-optical trap
(MOT) and transferred over 40 cm to our science cell. Differential pumping between chambers ensures science cell pressures of $<3\times
10^{-11}$ torr, almost two orders of magnitude lower than the pressure in the MOT chamber. The atoms are then further cooled by rf evaporation
to form a BEC with roughly $5\times10^{5}$ atoms in an electronic ground state $|F=1,m_{F}=-1>$. We create the cigar-shaped condensate in a
Ioffe-Pritchard magnetic trap with trapping frequencies of 230 and 6.4 Hz in the radial and axial directions, respectively, resulting in
Thomas-Fermi radii of 3.05 and 110$~\mu$m. Inside the science cell, a monolithic pyrex table holds the substrates that were studied, as shown in
Fig.\ref{fig:cell}. The surface used for the majority of this experiment was an 870 nm thick layer of yttrium that was deposited onto a polished
fused silica substrate by means of electron-gun vapor deposition. This coating technique, as opposed to the use of a bulk piece of polished
yttrium, ensures a surface free of contamination from the polishing process. However, atomic force microscope (AFM) scans of the surface
revealed structural irregularities with a grain size of $\sim$ 25 nm and a peak-to-valley range of $\sim$9 nm.  A more detailed description of
our apparatus, including our surface-distance calibration techniques, can be found in~\cite{jeffPRA,davePRA}.

\section{Results and Discussion}

\subsection{Electric field detection}

Since the surface ideally emits fairly weak electric fields, we create measurable fields by purposely depositing ultracold rubidium atoms onto
the metal layer, as described in~\cite{jeffPRA}. Briefly, a rubidium atom adsorbed onto the surface changes its atomic level structure in such a
way that its valence electron partially resides inside the metal. The resulting charge separation ($\sim1$\AA) effectively creates a dipole
aligned normal to the surface; the dipole's strength is related to the electronegativities of the involved substances.  To minimize this effect
for our studies of atom-surface interactions, we chose a metal with a low work function for our surface~\cite{wf}.

Although the motion of neutral atoms is insensitive to uniform electric fields, field gradients will create forces that cause significant
perturbations to the atoms' trapping potential. Even during ideal operations, it is unavoidable to deposit rubidium atoms onto the surface;
these atoms produce small, uniform field gradients. Since this type of electric field is one of the major systematics in precision surface-force
experiments~\cite{jeffPRA,davePRA,ourPRL}, purposely depositing atoms gives us the best tool to account for such errors.

When depositing atoms, we magnetically push low-density noncondensed atom clouds with dimensions ($\sim10 \mu$m radially) larger than our BEC
dimensions into the surface. The larger spatial extent of the deposited atoms provides more uniform field gradients across the cloud. Immediate
analysis of the resultant electric field shows that significant desorption or diffusion of adatoms at room temperature does not occur on
timescales of minutes, but rather several days. Atom diffusion and desorption will be discussed further in Sec.~\ref{sec:Heat}.

Our method of measuring electric field gradients, partially described in \cite{davePRA}, involves the application of an electric field via two
conducting plates mounted above and below the science cell, as shown in Fig.~\ref{fig:cell}. The plates consist of a thin layer of transparent
indium-tin-oxide (ITO) on $1\times10\times35$ mm$^{3}$ glass plates electrically connected to leads with a conductive epoxy. The use of ITO
allows optical access to the cell from the vertical direction, which is necessary for our laser heating method \cite{ourPRL}, and also leaves
open the possibility of imaging through the plates.

Previous studies have shown that an oscillating external electric field will drive a dipole oscillation of trapped neutral atoms if an electric
field gradient is present~\cite{davePRA,ourPRL}. An atom in an external electric field experiences an energy shift equal to
$U_{E}=-(\alpha_{o}/2)|\vec{E}|^{2}$, where $\alpha_{o}$ is the ground state static polarizability, and a force $\vec{F}$ equal to
$-\vec{\nabla}U_{E}$. Spurious forces that must be measured and accounted for to make claims of accuracy in precision surface-force measurements
therefore stem from field gradients:
\begin{equation}
\vec{F}(t)= \frac{\alpha_{o}}{2}\vec{\nabla}|\vec{E}(t)|^{2}.
\end{equation}

\begin{figure}
\leavevmode \epsfxsize=3.3in \epsffile{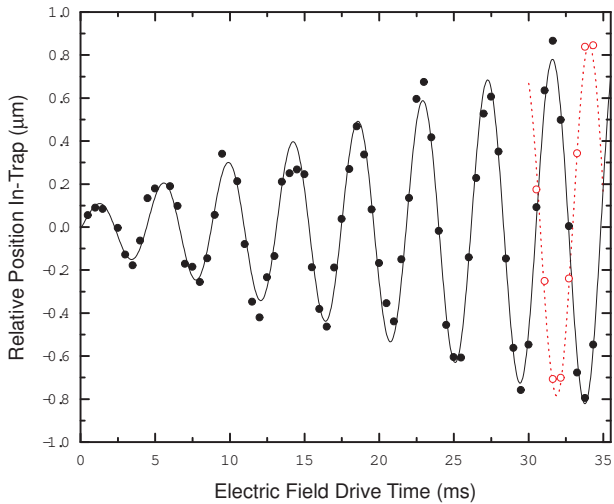}\caption{\label{fig:flipdata} (Color online). Data showing the relative position of
atoms in a resonantly driven mechanical oscillation (closed black points) with a fit to the data.  The rate of growth of the oscillation is
directly proportional to the electric-field gradient from surface contaminants.  The ability to determine the sign of the field gradient can be
seen when the polarity of the applied electric field is switched.  This corresponds to the open (red) circles, which show a clear $\pi$-phase
change in the oscillation.}
\end{figure}

If an external ac electric field is applied at the trap frequency $\omega_{o}$, then the system will act as a high-$Q$ resonantly driven
oscillator. The electric field from surface contaminants $\vec{E}^{*}$ and the applied external field
$\vec{E}^{ext}(t)=E^{ext}_{x}\cos(\omega_{o}t)\hat{x}$ act in tandem to resonantly drive the trapped atoms' motion with a time-varying force,
\begin{equation}
\vec{F}(t)= \frac{\alpha_{o}}{2}\vec{\nabla}(\vec{E}^{*}+\vec{E}^{ext}(t))^{2}.
\end{equation}

If one assumes that the applied electric field is much greater than the field to be measured ($E^{ext}_{x}>>E^{*}_{i}$) and invoking
$\vec{\nabla} \times \vec{E^{*}} \simeq 0$, the total forces on the atoms can be written as
\begin{equation}
F_{i}(t) \simeq \alpha_{o} E^{ext}_{x} \cos(\omega_{o} t)\partial_{x} E_{i}^{*},
\end{equation}
\noindent for $i = x,y,z$. The center-of-mass oscillation will then resonantly grow, as seen in Fig.~\ref{fig:flipdata}, as
\begin{equation}\label{eq:grow}
q_{i}(t) = \dot{a}_{i}~t~\cos(\omega t),
\end{equation}
\noindent where $q_{i}$ is the spatial coordinate in the $i$-direction.  The amplitude growth rate in the $i$-direction can then be expressed as
\begin{equation}\label{eq:Adot}
\dot{a}_{i} = \frac{\alpha_{o} E^{ext}_{x} \partial_{x} E^{*}_{i}}{2 m \omega_{o}},
\end{equation}
\noindent where $m$ is the mass of the atom.

We measure this growth rate by first transferring the atoms to an anti-trapped state and letting them expand for $\sim5$ ms.  Two
horizontal-imaging beams along $\hat{y}$ and $\hat{z}$ simultaneously image the atom cloud, which gives us information about the center-of-mass
position of the atom cloud in all three dimensions.  Thus, by measuring the resulting amplitude growth rate, we have a method to measure the
gradients of small electric fields which emanate from a surface. Fig.~\ref{fig:flipdata} shows the resulting oscillation of the resonantly
driven atom cloud (filled circles) in which the amplitude of the oscillation grows linearly with time.  As seen in Eq.(\ref{eq:Adot}), this
growth rate is proportional to the field gradient and becomes much smaller far from the surface or over a clean swath of surface, where field
gradients are small.

While Stark shifts are only sensitive to the magnitude of an electric field, our method can also determine the field gradient's direction. When
we drive the oscillating electric field, the oscillation begins with the field initially pointing in a known direction.  If the initial field
polarity is switched, however, the amplitude growth rate changes sign. For atoms starting from rest, the absolute value of the growth rate
remains unchanged, and the phase of the driven oscillation shifts by $\pi$, as shown by the open (red) circles in Fig.~\ref{fig:flipdata}. This
dependence on the phase of the applied electric field allows us to directly determine the direction of the field gradient in the $x$, $y$, and
$z$ directions at every measured point in space and thus to reconstruct the vector fields.

\begin{figure}
\leavevmode \epsfxsize=3.3in \epsffile{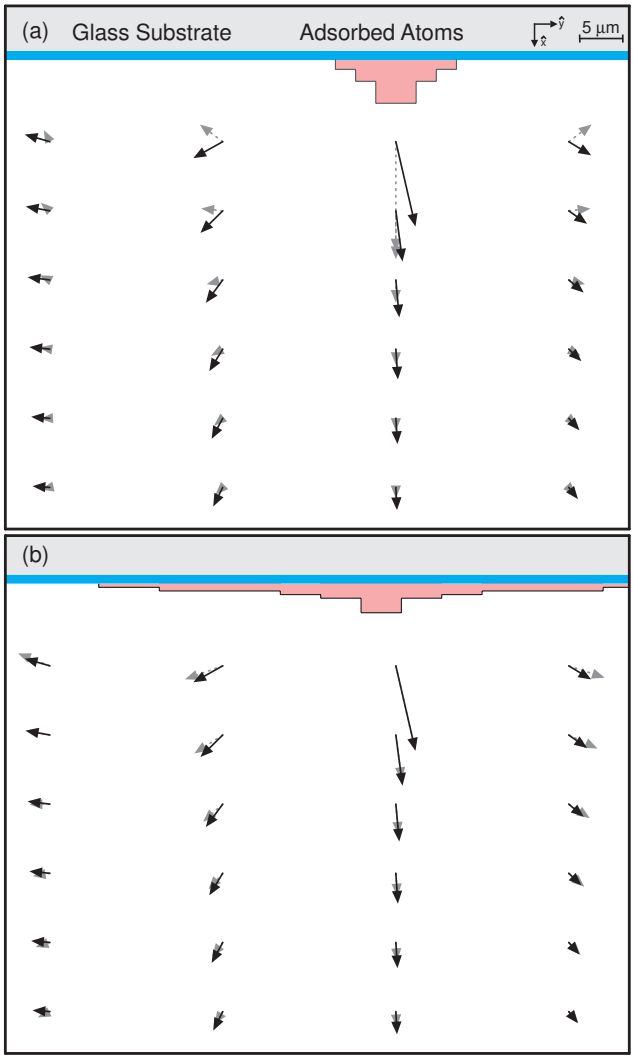}\caption{\label{fig:vectorfield} (Color online). Measurements of the electric-field
gradients at various positions along the yttrium surface let us reconstruct this 2-D vector plot (solid black arrows) of the electric field
generated by adsorbed rubidium atoms (pink layer-cake structure). Analysis was performed in which the measured field was fit to the field of (a)
a relatively localized pattern of dipoles (whose spread in the $\hat{y}$-direction is equal to the spread of atoms as they are initially
deposited) and (b) a spatially diffuse pattern of dipoles. The results of the fits are shown as dotted (gray) arrows. The layered structure in
(a) indicates the approximate location, spatial extent and surface density of atoms adsorbed onto the surface. The peak surface density of
adatoms is much less than one monolayer and would not form a structure extending from the surface. The height of the cake indicates the local
surface density of dipoles. The axial ($\hat{z}$) size of the applied atoms is a few hundred microns. Fields measured along this axis were
negligibly small and are not shown. In this figure, the longest vector represents a field of $\sim$19 V/cm.}
\end{figure}

If one assumes that the electric fields go to zero far from the surface [$\vec{E^{*}}(x=\infty)=0$], then we can extract the magnitude and
direction of the field by integrating a functional fit of the electric-field gradient from $\infty \rightarrow x$. If this process is repeated
for various locations along the surface ($y$-direction), we can then map out a two-dimensional vector plot of the electric field, as shown in
Fig.~\ref{fig:vectorfield}.  The solid black arrows in (a) and (b) show the reconstruction of the vector field following the adsorption of
$\sim$7$\times 10^{7}$ atoms onto the yttrium surface (thick blue line). The dashed gray arrows in (a) indicate the calculated electric field
from a thin line of dipoles oriented along $\hat{x}$, extending in and out of the page, whose surface-adsorbate density is represented by the
pink layer-cake structure.

For a dipole distribution similar in extent to the density distribution of the atoms as they were initially deposited, the expected field
disagrees significantly with the measured field in both direction and magnitude. However, if we allow for variability in the number, center
position, and spatial width $\sigma_{y}$ of the adsorbate pattern, we find qualitative and quantitative agreement with an electric field
produced by a similar number of adsorbates to that in (a), but spread more diffusely across the surface ($\sigma_{y}$ = 26 $\mu$m) than the
pattern of adsorbates initially placed onto the surface.  The results of a fitting routine are shown in Fig.~\ref{fig:vectorfield}(b), where the
more diffuse pattern of dipoles used to model the electric field is shown smeared across the surface.

\subsection{Estimating the dipole moment of a single adatom}

\begin{figure}
\leavevmode \epsfxsize=3.3in \epsffile{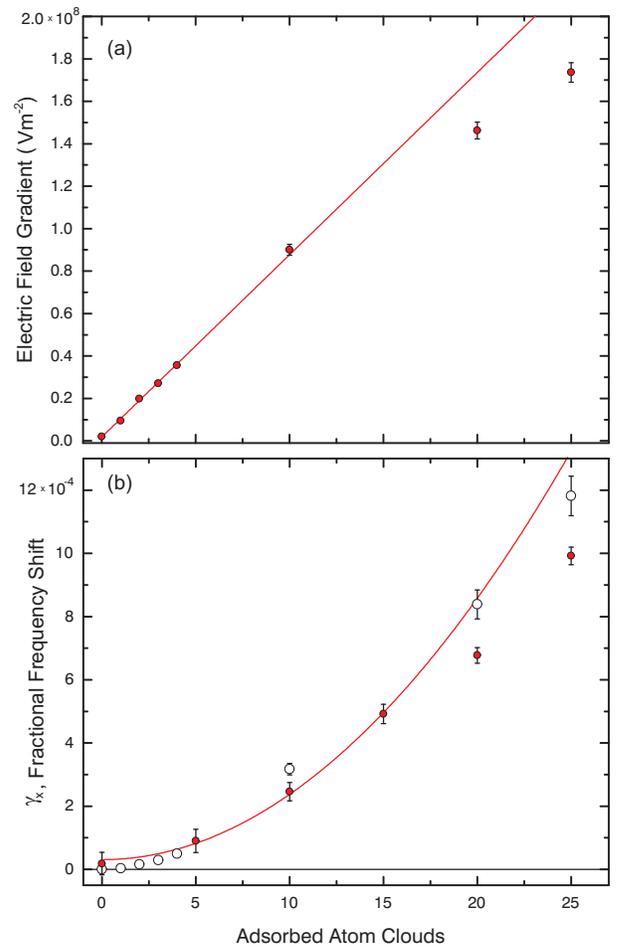}\caption{\label{fig:DvsN} (Color online). (a) Plot of the measured electric field gradient
versus number of clouds adsorbed onto an yttrium surface.  The linearity breaks down eventually, possibly because of surface effects such as
adatom-adatom interactions. (b) Measured fractional frequency shift data (closed red circles) with the expected fractional frequency shift (open
circles) obtained by processing the measured data in (a).  The solid lines (red) in (a) and (b) correspond to linear and quadratic fits,
respectively, for data in the 0--15 cloud regime. These fits illustrate the dependencies of each on the number of adsorbates
[Eqs.(\ref{eq:Adot}) and (\ref{eq:TFFS}), respectively].  The number of atoms in a single cloud is $\sim1.7 \times 10^{6}$.  This data was taken
at $x$ = 10 $\mu$m from the surface.}
\end{figure}

The precise characterization of the electric field lets us determine certain properties of the surface.  As mentioned earlier, the adsorption of
a rubidium atom onto a metal surface creates a surface dipole adsorbate whose strength depends upon the work function of the metal and the
ionization energy of the rubidium atom. Previous studies have shown that electric fields from these dipoles can be very large on metals, which
prompted the use of glass for our Casimir-Polder force studies~\cite{jeffPRA,davePRA}. To repeat Casimir-Polder experiments over metal surfaces,
metals with work functions lower than the ionization energy of the probe atom would be beneficial to study.  By carrying out our electric-field
studies, we can determine the dipole moment of an individual rubidium atom adsorbed onto yttrium and compare it to those from other surfaces.

To determine the dipole moment of a single adsorbed atom, we use a modelling program to match the measured field gradient with a calculated
field gradient. Our model creates a distribution of $N$ surface dipoles oriented normal to the surface; the physical parameters of the
distribution match those of the atom cloud. We then calculate the resulting fields and field gradients that emanate from the surface and compare
them with the measured values.  The only adjustable variable in this model is the dipole moment of one adsorbed atom, which is varied to match
calculated and measured field gradients. We neglect to add any surface diffusion process into the modelling program because the measurements to
which we compare were made rapidly with respect to surface diffusion times.

Fig.\ref{fig:DvsN}(a) shows the measured electric-field gradient versus the number of atom clouds adsorbed onto the surface. The linearity of
the measured field gradient (0--10 clouds) allows us to assume that the field generated from one adsorbed atom is identical to the field
generated by one cloud of atoms divided by the number of atoms in that cloud. From this, we find a relation between the electric field gradient
and the number of atoms deposited.  Using the procedure described above, we find that the dipole moment per Rb atom adsorbed onto our yttrium
surface is $\sim$35 Debye~\cite{notbulk} (corresponding to the valence electron residing within the substrate $\sim1$\AA from the Rb
center-of-mass, roughly one metallic bond length). We also measure a dipole moment of $\sim$3.2 Debye for Rb on fused silica, $\sim$5.4 Debye
for Rb on a metallic hafnium surface, and $\sim$19 Debye for Rb on a metallic lutetium surface.

\subsection{Accounting for systematic errors from electric fields}

With our knowledge of electric fields from surface contamination, we can accurately account for frequency shifts of dipole oscillations, like
those made in~\cite{jeffPRA,davePRA,ourPRL} that make precision measurements of surface forces.  The additional forces from surface contaminants
perturb the trapping potential near the surface in such a way that the perturbations result in an unwanted systematic shift of the data.  To
rule out this systematic shift, one may carefully measure field gradients from the surface and calculate the expected frequency shift as
follows:

Atoms trapped in a quadratic potential will see perturbations to the trapping frequency that are proportional to the curvature of the perturbing
potential,
\begin{equation}\label{eq:FS}
\Delta \gamma_{x} \approx \frac{-\partial_{x}^{2} U_{E}}{2 m \omega_{0}^{2}},
\end{equation}
\noindent where $\gamma_{x}$ is the change in trap frequency in the $x$-direction, normalized to the unperturbed trap frequency $\omega_{o}$,
\begin{equation}\label{eq:gamma}
\gamma_{x} = 1 - \frac{\omega_{x}}{\omega_{o}}.
\end{equation}
\noindent We can define $\Delta \gamma_{x}$ as the contribution to the fractional frequency shift due to the additional surface adsorbates,
$U_{E}=-(\alpha_{o}/2)|\vec{E^{*}}|^{2}$. Eq.(\ref{eq:FS}) then becomes
\begin{equation}\label{eq:FFS}
\Delta \gamma_{x} = \frac{\alpha_{o}}{2 m \omega_{o}^{2}} \sum_{i} ((\partial_{x}E_{i}^{*})^{2} + E_{i}^{*} \partial_{x}^{2}E_{i}^{*}).
\end{equation}

If we choose a convenient fitting form of the electric field that approximates the field generated by electrostatic patches, points, and lines
for a restricted range of $x$,
\begin{equation}\label{eq:Efield}
E_{i}^{*} = C_{i}x^{-p_{i}},
\end{equation}
\noindent Eq.(\ref{eq:FFS}) then becomes,
\begin{equation}\label{eq:TFFS}
\Delta \gamma_{x} = \frac{\alpha_{o}}{2 m \omega_{o}^{2}} \sum_{i} (2 p_{i}+1) p_{i} C_{i}^{2} x^{-2 (p_{i}+1)}.
\end{equation}
\noindent We can then extract $C_{i}$ and $p_{i}$ from measurements of the amplitude growth rate $\dot{a}_{i}$ at various displacements from the
surface. The $C_{i}$ coefficients can then be written as
\begin{equation}
C_{i} = \frac{2 m \omega_{i} \dot{a}_{i}}{\alpha_{o} E^{ext}_{x} p_{i}},
\end{equation}
\noindent where $\omega_{i}$, the frequency of the applied electric field, is chosen to be the trap frequency in the $i$-direction.

As shown in Fig.~\ref{fig:DvsN}(a), the measured electric field gradient increased linearly with the number of adsorbed atoms. If we assume that
the power-law dependence of the electric field does not change significantly with the number of applied atoms, we can then deduce that the
coefficient $C_{i}$ is proportional to number of applied atoms.  This implies, from Eq.(\ref{eq:TFFS}), that the trap frequency a fixed distance
from the surface will vary quadratically with the number of adsorbates as well, since it is proportional to $C_{i}^{2}$. Fig.\ref{fig:DvsN}(b)
shows data verifying that indeed the fractional change in trap frequency from the adsorbates $\Delta \gamma_{x}$ varies quadratically with the
number of adsorbed atoms.  The open circles in (b) show the results of the above analysis on the data in (a) and agree well with measured
values.

\begin{figure}
\leavevmode \epsfxsize=3.3in \epsffile{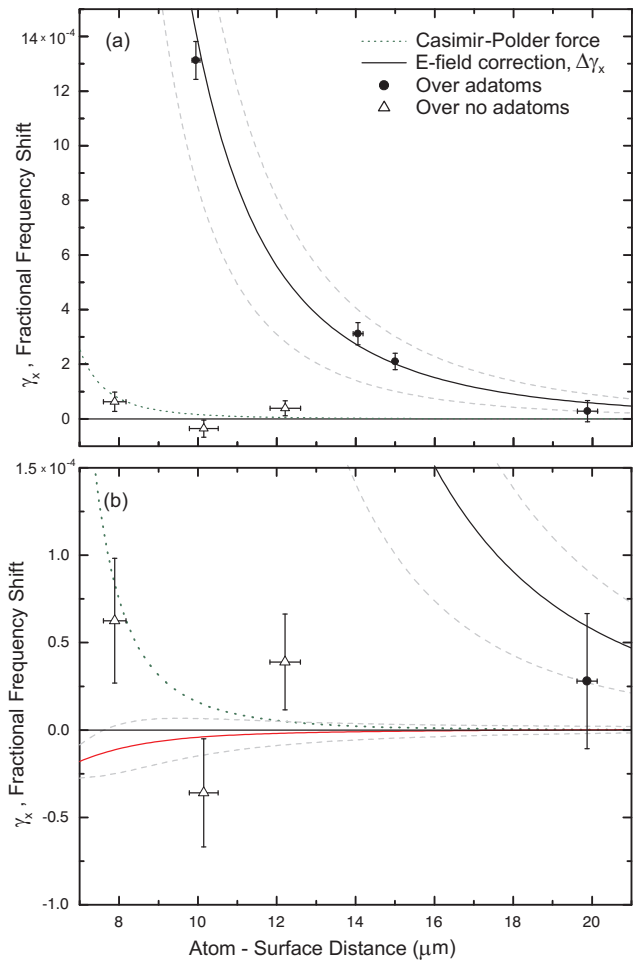}\caption{\label{fig:FFS} (Color online). (a) The measured fractional frequency shift for two
separate surface locations.  The filled circles are data taken directly over 7 $\times 10^{7}$ adsorbed atoms, and the solid black line is the
shift predicted from electric-field measurements. The dashed gray lines represent error bounds obtained from field-gradient measurements. The
open triangles are data taken over a `clean' surface location (no adatoms) where the electric-field correction is consistent with zero (solid
red line in (b)). The correction to data taken over a clean spot is frequently small enough to exhibit a two-component power law dependence for
which the correction may, in fact, be slightly negative. The expected Casimir-Polder shift is shown with a green dotted line.}
\end{figure}

With these calculations in hand, we can accurately predict the systematic fractional frequency shift by directly measuring the electric field
gradient. The results in Fig.~\ref{fig:FFS} show the fractional frequency shift as a function of distance to the surface for two separate
locations on the surface. The open triangles were taken over a clean area, where we measured a negligible electric field; the filled circles
were taken over a surface location in which we purposely adsorbed $\sim7 \times 10^{7}$ atoms.  The solid black line represents the theoretical
fractional frequency shift predicted by Eq.(\ref{eq:TFFS}), corresponding to measurements made of the electric field emanating from that surface
location. The agreement between data and theory illustrates that we can accurately account for frequency shifts from electric fields. For the
purpose of characterizing systematic errors to surface-force measurements, our method of characterizing the surface quality of the patch of
surface in which small force measurements are made is more directly relevant than canonical surface-science techniques that involve AFM and
scanning electron-microscope surface imagery.

\subsection{Diffusing adsorbates with heat} \label{sec:Heat}

\begin{figure}
\leavevmode \epsfxsize=3.3in \epsffile{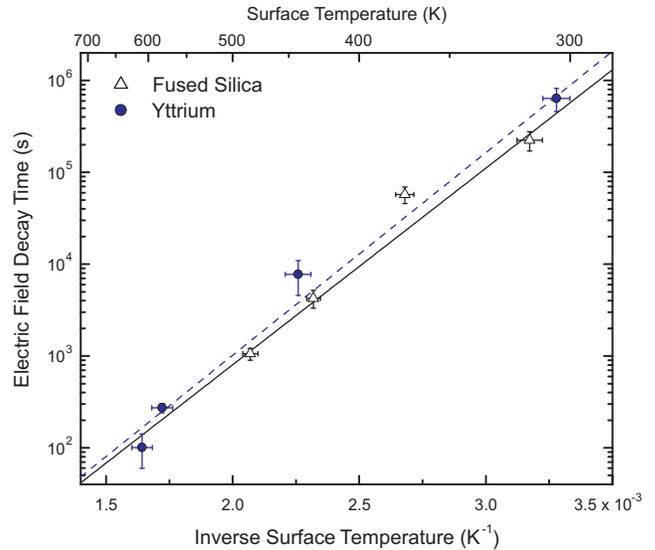}\caption{(Color online) Decay time of the stray electric field as a function of temperature
for a fused silica substrate (open triangles) and an yttrium surface (blue filled circles).}\label{fig:desorb}
\end{figure}

Apart from characterizing the atoms on a surface and their resulting electric fields, one might also like to demonstrate a way to lessen the
strength of the electric fields and their undesirable effects. We achieve this by applying heat to our substrate via a high-power laser. The
back surface of the fused silica substrate is coated with a $\sim100\mu$m thick layer of graphite, which is opaque to an infrared heating laser
($\lambda =$ 860 nm). Heating the surface provides enough thermal energy for surface contaminants to redistribute themselves across the surface
or to desorb entirely.  The temperature of the metallic surface is calibrated versus the power of the heating laser using the same methods as
in~\cite{ourPRL}.  This technique should not be confused with light-induced atomic desorption, in which adatoms absorb ultraviolet light and
desorb from the surface; in our case, no laser light directly impinges on the adatoms or on the surface to which they are attached.

The exponential decay time of an electric field emanating from the surface can come from either a desorption process, in which the adatoms
escape from the surface-binding potential, or a diffusion process, in which the atoms overcome a smaller hopping energy and hop from site to
site, redistributing themselves across the surface.  The time scale $\tau$ for desorption and diffusion events to take place is characterized by
the temperature of the surface $T$, the energy of activation $E_{A}$, and an attempt rate $\gamma_{o}$ that depends upon the surface process,
\begin{equation}
\tau(T) = \gamma_{o}^{-1}e^{E_{A}/kT}.
\end{equation}

The results of this and a similar study over a fused silica surface can be seen in Fig.~\ref{fig:desorb}. The similar fits to the data suggest
that rubidium has similar activation energies on fused silica and yttrium ($E_{A}\approx$ 0.42 eV on each) and also reveal $\gamma_{o}$ to be
approximately 15--25 s$^{-1}$. This measured attempt rate is $\sim$10 orders of magnitude smaller than what one would expect for a desorption
process and seems more characteristic of a surface diffusion process that results from numerous random-walk hops. Fig.~\ref{fig:desorb} also
shows that it is indeed possible to lessen the undesired systematic effects of electric fields on surfaces by baking.

\section{Conclusion}

We have elaborated on a technique that uses magnetically trapped neutral atoms as a tool for measuring small electric-field gradients and their
effects. We use trapped atoms in a high-$Q$, driven harmonic oscillator for our measurements, such that small gradients on the order of
$\sim$300 nV/$\mu$m$^{2}$ are measurable. Our technique allows us to reconstruct the full electric vector field from surface contaminants. Using
our techniques, we are able to analyze and account for systematic errors in precision surface-force measurements that use mechanical dipole
oscillations to measure small surface forces. Electric fields were also shown to decay significantly when heat was applied to the substrate.
This baking technique aided in removing unwanted systematic effects associated with surface contaminants.

\section*{ACKNOWLEDGEMENTS}

We are grateful for surface preparations done by D.~Alchenberger and useful conversations with the JILA ultracold atoms and molecules
collaboration. This work was supported by grants from the NSF and NIST.


\begin{thebibliography}{10}
\bibitem{wf}
The work functions of Y, Hf, and Lu are 3.1, 3.9, and 3.3 eV, respectively, all of which are lower than the ionization energy of rubidium (4.2
eV). All three metals are stable in air and vacuum as well as relatively nonmagnetic.
\bibitem{notbulk}
The surface structure that results from our yttrium deposition techniques could cause adsorption dipoles to be different from dipoles on bulk
yttrium.
\end{thebibliography}
\end{document}